\documentclass[epj]{svjour}
\usepackage{graphics}
\usepackage{epsfig}
\def\n {\nonumber}

\def\M  {{\cal M}}
\def\F  {{\cal F}}
\def\O  {{\cal O}}
\def\R  {{\cal R}}

\def\J {$J/\psi$ }
\def\J2 {$\psi(2S)$}

\def\j { J/\psi  }

\def\ktf {$k_t$-factorization }

 \newcommand\beq{\begin{equation}}
 
 \newcommand\eeq{\end{equation}}
 \newcommand\beqn{\begin{eqnarray}}
 \newcommand\eeqn{\end{eqnarray}}

\def\cpc#1#2#3  {{Computer\ Phys.\ Comm.\ }  {\bf#1}, (#3) #2}
\def\chpc#1#2#3 {{Chin.\ Phys.\ C }          {\bf#1}, (#3) #2}
\def\err#1#2#3  {{\it Erratum }              {\bf#1}, (#3) #2}
\def\epjc#1#2#3 {{Eur.\ Phys.\ J.\ C }      {\bf#1}, (#3) #2}
\def\epjp#1#2#3 {{Eur.\ Phys.\ J.\ Plus }    {\bf#1}, (#3) #2}
\def\dum#1#2#3  {{~}                         {\bf#1}, (#3) #2}
\def\ib#1#2#3   {{\it ibid. }                {\bf#1}, (#3) #2}
\def\jcp#1#2#3  {{J.\ Comput.\ Phys.\ }      {\bf#1}, (#3) #2}
\def\jetpl#1#2#3 {{\rm JETP Lett. }          {\bf#1}, (#3) #2}
\def\jhep#1#2#3 {{JHEP }                     {\bf#1}, (#3) #2}
\def\ijmp#1#2#3 {{Int.\ J.\ Mod.\ Phys.\ }   {\bf#1}, (#3) #2}
\def\jpg#1#2#3  {{J.\ Phys.\ G }             {\bf#1}, (#3) #2}
\def\mpl#1#2#3  {{Mod.\ Phys.\ Lett.\ }      {\bf#1}, (#3) #2}
\def\mpla#1#2#3 {{Mod.\ Phys.\ Lett.\ A }    {\bf#1}, (#3) #2}
\def\ncim#1#2#3 {{Nuovo Cimento }            {\bf#1}, (#3) #2}
\def\np#1#2#3   {{Nucl.\ Phys.\ }            {\bf#1}, (#3) #2}
\def\npb#1#2#3  {{Nucl.\ Phys.\ B }          {\bf#1}, (#3) #2}
\def\pan#1#2#3  {{Phys.\ At.\ Nuclei }       {\bf#1}, (#3) #2}
\def\plb#1#2#3  {{Phys.\ Lett.\ B }          {\bf#1}, (#3) #2}
\def\prep#1#2#3 {{Phys.\ Rep.\ }             {\bf#1}, (#3) #2}
\def\prd#1#2#3  {{Phys.\ Rev.\ D }           {\bf#1}, (#3) #2}
\def\prl#1#2#3  {{Phys.\ Rev.\ Lett.\ }      {\bf#1}, (#3) #2}
\def\ptp#1#2#3  {{Prog.\ Theor.\ Phys.\ }    {\bf#1}, (#3) #2}
\def\ps#1#2#3   {{Physica Scripta }          {\bf#1}, (#3) #2}
\def\rmp#1#2#3  {{Rev.\ Mod.\ Phys.\ }       {\bf#1}, (#3) #2}
\def\rpp#1#2#3  {{Rep.\ Prog.\ Phys.\ }     {\bf#1}, (#3) #2}
\def\sa#1#2#3   {{Sci.\ Acta}                {\bf#1}, (#3) #2}
\def\sjnp#1#2#3 {{Sov.\ J.\ Nucl.\ Phys.\ }  {\bf#1}, (#3) #2}
\def\spj#1#2#3  {{Sov.\ Phys.\ JETP }        {\bf#1}, (#3) #2}
\def\spjl#1#2#3 {{Sov.\ JETP Lett.\ }        {\bf#1}, (#3) #2}
\def\spu#1#2#3  {{Sov.\ Phys.-Usp.\ }       {\bf#1}, (#3) #2}
\def\yaf#1#2#3  {{Yad.\ Fiz.\ }              {\bf#1}, (#3) #2}
\def\zp#1#2#3   {{Zeit.\ Phys.\ }            {\bf#1}, (#3) #2}
\def\zpc#1#2#3  {{Z.\ Phys.\ C }             {\bf#1}, (#3) #2}
\def\etal {{\it et al. }}

\begin{document}

\title{Fragmentation of charm to charmonium\\ in \boldmath$e^+e^-$ and $pp$ collisions }
\author{S.\ P.\ Baranov\inst{1} \and B.\ Z.\ Kopeliovich\inst{2}}
\institute{P.N. Lebedev Institute of Physics, 
              Lenin Avenue 53, 119991 Moscow, Russia \and Departamento de F\'{\i}sica,
Universidad T\'ecnica Federico Santa Mar\'{\i}a; and
Centro Cient\'ifico-Tecnol\'ogico de Valpara\'iso;
Avenida Espa\~na 1680, Valpara\'iso, Chile}
\date{\today}
\abstract{
We perform numerical comparison of the fragmentation mechanism of charmonium production  
($g\,g\to c\,\bar{c}$ followed by $c\to\psi\,c$) with the full leading order 
calculation ($g\,g\to\psi\,c\,\bar{c}$ at $\O (\alpha_s^4)$). 
We conclude that the non-fragmentation contributions remain important up 
to $J/\psi$ transverse momenta about as large as 40 GeV, thus making questionable the 
applicability of the fragmentation approximation at smaller transverse momenta.
} 
\maketitle
%

\section{Introduction}

The general 
Factorization principle and the concept of quark and gluon fragmentation 
functions \cite{ColSop} constitute a widely exploited framework to describe 
particle production phenomena at collider energies.
The method is proved to be mathematically consistent in the region of
asymptotically high transverse momenta of the produced particles.

The goal of the present consideration is to examine the universality of the 
quark fragmentation function and to outline the kinematic conditions when 
the fragmentation approach can be trusted as a reliable approximation.
Our present study was to some extent triggered by the paper \cite{Bodwin},
where the fragmentation approach was used to describe the experimental
data (CDF, ATLAS, CMS) at $p_{\psi T}>10$ GeV.

To carry out this task, we make a comparison of two calculations. 
First, we consider an $\O (\alpha_s^2)$ subprocess $g\,g\to c\,\bar{c}$ and
convolute it with an $\O (\alpha_s^2)$ fragmentation function $c\to\psi\,c$,
where $\psi$ is meant to be either  $\j$ or $\psi(2S)$.
Second, we perform a full $\O (\alpha_s^4)$ calculation for the process
$g\,g\to\psi\,c\,\bar{c}$ and see to what extent does the `full result' 
matches the fragmentation interpretation.

\section{Perturbative 
color-singlet 
fragmentation \boldmath$c\to\psi\bar cc$ }

To calculate the charmed quark fragmentation function, we start with the
process $e^+e^-\to\gamma^*\to\psi\,c\,\bar{c}$ considered in the virtual 
photon rest frame with the $z$ axis oriented along the negative direction 
of the charmed antiquark momentum. The corresponding Feynman diagrams are 
displayed in Fig.~\ref{fig:annihilation}. 
\begin{figure}
\centering
\epsfig{figure=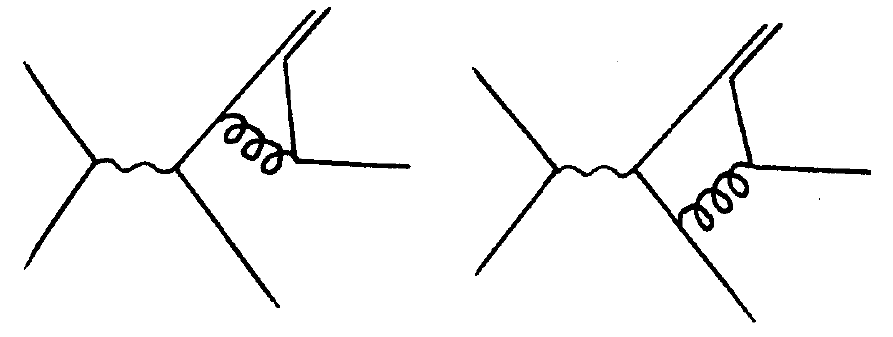,width=6cm}
\caption{\label{fig:annihilation} Feynman diagrams used to calculate the $c\to\psi$ fragmentation
function from $e^+e^-$ annihilation, $e^+e^-\to\gamma^*\to\psi+c+\bar{c}$.}
\label{fig:annihilation}
\end{figure}

The fully differential cross section then reads
\begin{eqnarray}
d\sigma &=& \frac{1}{2s}\,\frac{1}{(2\pi)^5}\,   \label{dsigma}
            |\M (ee\to\gamma^*)|^2\,\cdot\,
            |\M (\gamma^*\to\psi\,c\,\bar{c})|^2\, \n \\ 
 ~ &\times&  \frac{1}{M(\psi\,c\,\bar{c})^4}\,          
            \frac{\lambda^{1/2}(s,p^2,m_c^2)}{8s}\, 
            \frac{\lambda^{1/2}(p^2,m_\psi^2,m_c^2)}{8p^2}\n \\[2mm]
 ~ &\times& d\Omega\;dp^2\,d\phi\;d\cos\theta,
\end{eqnarray}
where $s$ is the overall invariant energy; $p_\psi$, $p_1$ and $p_2$ the 
4-momenta of $J/\psi$ meson and the charmed quark and antiquark, respectively;
$\Omega$, $\phi$, and $\theta$ the angular variables of the reaction;
$\lambda$ is the standard `triangle' kinematic function \cite{BycKaj};
and the momentum $p=p_1+p_\psi$ represents the fragmenting (or `parent')
quark momentum.

The above formula can be interpreted as a product of the quark production 
cross section
\begin{eqnarray}
&& \!\!\!\!\!\!\!\!\!\! d\sigma(e^+e^-\to c\,\bar{c}) = \n \\
&& \frac{1}{2s}\,\frac{1}{(2\pi)^2}\, \label{ee2cc}
   \frac{\lambda^{1/2}(s,p^2,m_c^2)}{8s}\;
   |\M (ee\to c\,\bar{c})|^2\;d\Omega
\end{eqnarray}
and the $c$-quark fragmentation probability. 
After dividing Eq.(\ref{dsigma}) by Eq.(\ref{ee2cc}) 
we arrive at the definition of the differential fragmentation function

\begin{eqnarray}
&& \!\!\!\!\! dD(c^*\to\psi\,c) \label{dD} = 
\label{Dz}\\
&& \!\!\!\! \frac{1}{(2\pi)^3}
  \frac{|\M (\gamma^*\to\psi\,c\,\bar{c})|^2}{|\M (\gamma^*\to c\,\bar{c})|^2}\,
\lambda^{1/2}(p^2,m_\psi^2,m_c^2)\;dp^2\,d\phi\;d\cos\theta.\n
\end{eqnarray}

The latter can be further reduced to the conventional fragmentation
function $D_{c/\psi}(z)$ by introducing the light-cone variable 
$z=p^+_\psi/p^+=(E_\psi+p_{\psi,z})/(E+p_{z})$ and integrating over all 
other variables in Eq.(\ref{dD}):
\begin{equation}
D_{c/\psi}(z)=\! \int\!\! D(c^*\to\psi\,c)\delta(z - p^+_\psi/p^+)
   dp^2 d\phi \,d\cos\theta.
\end{equation}

\begin{figure}
\centering
\epsfig{figure=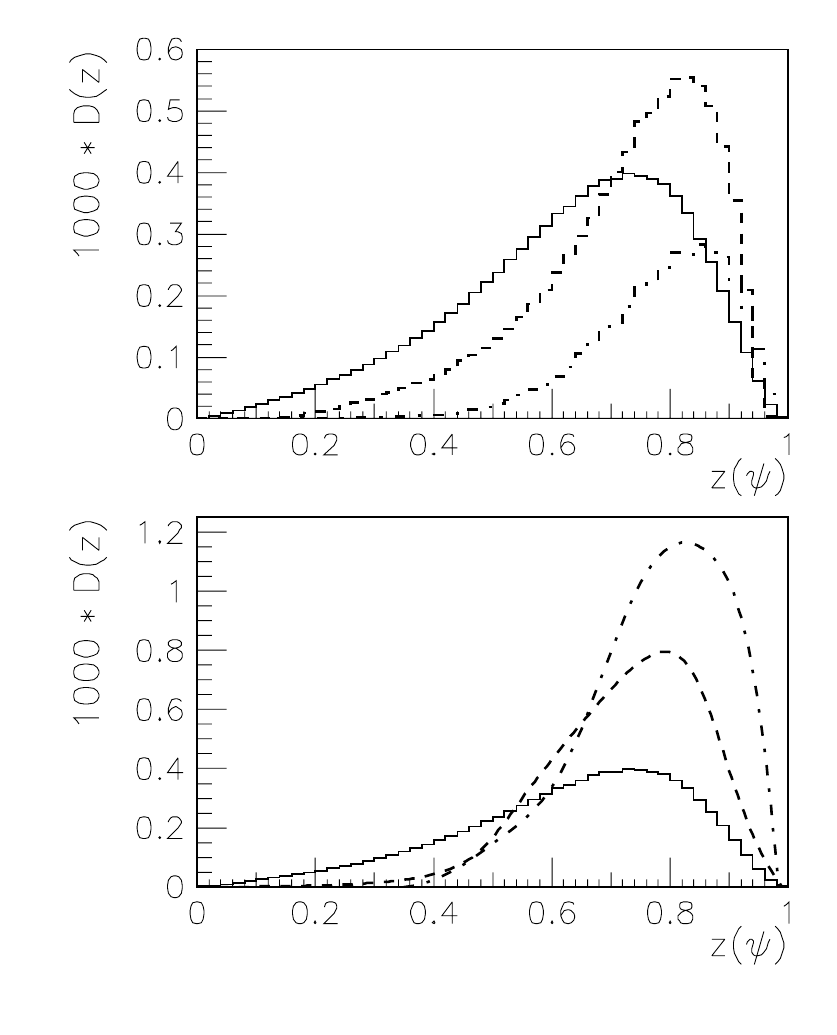,width=8.5cm}\\[-0.8cm]
\caption{\label{fig:Cfragm} Effective $c\to\psi$ fragmentation functions derived from
different partonic subprocesses:
solid curve, $e^+e^-\to\gamma^*\to\psi c\bar{c}$, calculated with Eq.~(\ref{Dz}).
Other curves are calculated for $g\,g\to\psi\,c\,\bar{c}$ as is described in Sect.~\ref{pp}.
Dashed curves  for $p_{\psi T}>20$ GeV, $p^*_{T}>20$ GeV;
dash-dotted curves for  $p_{\psi T}>50$ GeV, $p^*_{T}>50$ GeV.
Upper plot, sorrsponds to the collinear scheme with MSTW gluon densities \cite{MSTW08}; 
lower plot, \ktf with A0 gluon densities \cite{Jung}.}
\end{figure}

Calculations show almost no dependence on $e^+e^-$ energy,
what confirms the full dominance of the fragmentation regime.
Our results are plotted in Fig.~\ref{fig:Cfragm} by solid curves. They are 
fully consistent with other $\O (\alpha_s^2)$ calculations 
presented in the literature \cite{Zhang,Chung,Cheung}.
Comparison with data would need inclusion of higher order corrections,
can be done effectively in terms of radiational energy loss \cite{jet-lag},
however we prefer here to stay with LO order approximation, aiming at the
comparison with other LO results.

\section{Proton-proton collision}\label{pp}
\subsection{Glue-glue fusion}

\begin{figure}
\centering
\epsfig{figure=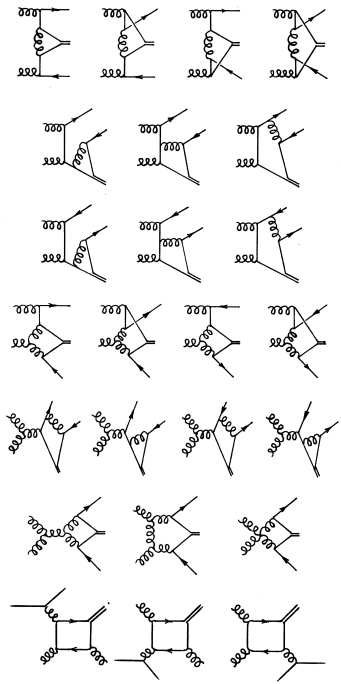,width=7cm}
\caption{\label{fig:fusion} Feynman diagrams representing the gluon-gluon fusion process 
$gg\to\j+c+\bar{c}$ at full leading order. Only a few of the listed diagrams 
can be interpreted as a $c$-quark fragmentation.}
\end{figure}

In $pp$ collisions the leading order process of $\j\, \bar cc$ production is glue-glue fusion,
\begin{equation}
      g+g\to J/\psi + c + \bar{c}. \label{gluglu}
\end{equation}
We employ the Feynman diagrams depicted in Fig.~\ref{fig:fusion}, which are all necessary 
to compose a gauge invariant set (for more details see \cite{myJcc},
where one can find explicit algebraic expressions for all of these diagrams).

In fact, we will perform two calculations in parallel, using the odinary 
(collinear) and the \ktf approaches. The latter can be treated as an 
effective Next-to-Leading order (NLO) calculation, since a significant part 
of higher-order radiative corrections is  taken into account in the 
form of $k_T$-dependent (unintegrated) gluon densities.

The evaluation of Feynman diagrams is straightforward and follows the
standard QCD rules. For the initial off-shell gluons (if any) 
If we adopt 
the \ktf ~prescription \cite{GLR83} for the initial off-shell gluons, the spin density matrix 
is taken in 
the form
$\overline{\epsilon_g^{\mu}\epsilon_g^{*\nu}}=k_T^\mu k_T^\nu/|k_T|^2,$
where $k_T$ is the component of the gluon momentum normal to the
beam axis. 
In the collinear limit, when $k_T\to 0$, this expression converges to the 
ordinary $\overline{\epsilon_g^{\mu}\epsilon_g^{*\nu}}=-g^{\mu\nu}/2$,   
while in the case of off-shell gluons it contains an admixture of
longitudinal polarization.
Calculation of the traces and of all Feynman 
diagrams was done with the algebraic system {\sc form} \cite{FORM}.

To purify the theoretical analysis, we will restrict it to a comparison between 
the color-singlet \cite{Chang,Baier,Berger} calculations only; that would
is sufficient for the goal of the present study.
The produced color-singlet $\bar cc$ dipole of a small separation $\sim 1/m_c$,
should be projected to the large-size charmonium wave function. Neglecting the small size of the dipole, one arrives at 
a simple result, the value of the radial wave function at the origin  
$|{\R}(0)|^2$. While the charmonium wave function with realistic potentials is known, and 
its convolution with the dipole size distribution would be more accurate \cite{kz91,hikt},
we rely here on the small-dipole approximation.
Then, the $\psi$ production probability contains only one parameter,   
$|{\R}(0)|^2$, which is known from the charmonium leptonic decay width 
\cite{PDG}. 

Summarizing, the fully differential cross section reads 
\begin{eqnarray}
&& \!\!\!\!\!\! d\sigma(pp\to\psi c\bar{c}X)= \n \\
&& \frac{\pi\alpha_s^4}{3\hat{s}^2}\,\frac{|{\R}(0)|^2}{4\pi}
   \frac{1}{4}\sum_{\mbox{{\tiny spins}}}\;
   \frac{1}{64}\sum_{\mbox{{\tiny colors}}}
   |{\cal M}(gg\to\psi c\bar{c})|^2 \n \\
&& \times\, {\F}_g(x_1,k_{1T}^2,\mu^2)\;{\F}_g(x_2,k_{2T}^2,\mu^2)\n \\[1mm]
&& \times\, dk_{1T}^2\,dk_{2T}^2\,dp_{\psi T}^2\,dp_{cT}^2\,
   dy_\psi\,dy_c\,dy_{\bar{c}}\n \\[1mm]
&& \times\, \frac{d\phi_1}{2\pi}\,\frac{d\phi_2}{2\pi}\,
   \frac{d\phi_\psi}{2\pi}\,\frac{d\phi_c}{2\pi}, \label{lips}
\end{eqnarray}
where $s$ is the total initial invariant energy squared,
$\hat{s}$ the squared energy of the partonic subprocess, 
$x_1$ and $x_2$ the parton light-cone momentum fractions;
$k_{1T}$, $k_{2T}$, $\phi_1$ and $\phi_2$ the transverse momenta 
and azimuthal angles of the initial (off-shell) gluons, and
$y_\psi$, $y_c$, $y_{\bar{c}}$, $p_{\psi T}$, $p_{cT}$, $p_{\bar{c}T}$,
$\phi_\psi$, $\phi_c$ and $\phi_{\bar{c}}$
the rapidities, transverse momenta and azimuthal angles of $\psi$ 
and the accompanying charmed quark and antiquark, respectively.

Throughout this paper, we use the "A0" parametrization \cite{Jung,CASCADE}
for the $k_T$-dependent gluon densities ${\F}_g(x_i,k_{iT}^2,\mu^2)$, 
with $\mu^2=\hat{s}/4$.
For collinear calculations we omit the integration over $k_{1T}$ and $k_{2T}$ 
and use the MSTW leading-order set \cite{MSTW08} for the ordinary gluon 
distribution functions.
The multidimensional integration in (\ref{lips}) has been performed by
means of the Monte-Carlo technique, using the routine VEGAS \cite{VEGAS}.

\subsection{Theoretical experiment: "jet" reconstruction}

The results of calculation of the full set of graphs should contain 
a contribution of charm fragmentation. To quantify such a contribution one
needs to reconstruct the fragmenting quark
momentum. A  $\psi$ produced  this way should be accompanied with either
$c$ or $\bar{c}$, thus referring to the quark or antiquark fragmentation. 
To search for such correlations one can select the configurations of $\psi\bar cc$
with lowest two-body invariant mass, either ($M(\psi c) < M(\psi\bar{c})$),
or vice versa. Of course any of such correlations are disturbed by other hadrons produced from the debris 
of the colliding protons. In the case of collinear factorization the transverse momenta of such hadrons are limited and their influence should fade away at large $p_{\psi T}$. However, in the case of the approach, based on $k_T$ factorization, the tail of the gluon distribution, $1/k_T^4$, is close to the $p_T$ dependence of charm production. Therefore, in this case one cannot separate well the contribution of the diagrams in Fig.~\ref{gluglu} from the hadronic background, even at high $p_{\psi T}$.

\begin{figure}
\epsfig{figure=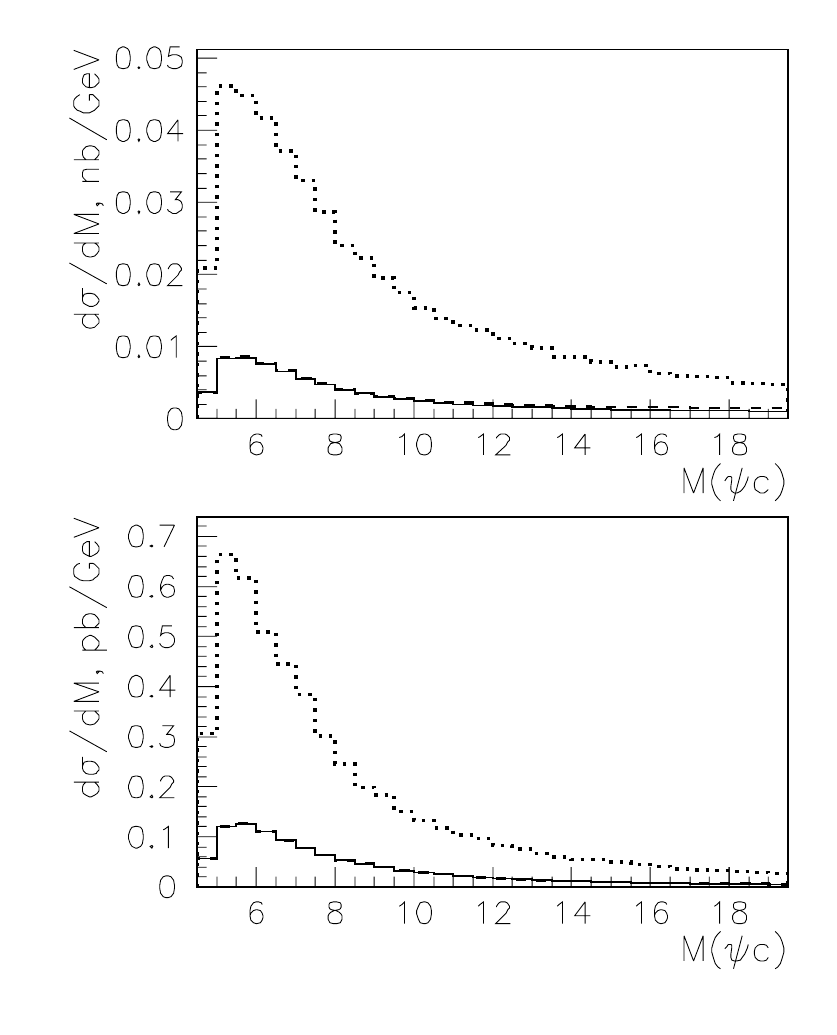,width=8.5cm}\\[-0.8cm]
\caption{\label{fig:mass} Invariant mass of the $J/\psi+c$ system as seen
under the different kinematic selection rules.
{\it Upper panel:}
solid curve, $p_{\psi T}>20$ GeV, $p^*_{T}>20$ GeV;
dashed curve, $p_{\psi T}>20$ GeV, $p^*_{T}>5$ GeV;
dotted curve, $p_{\psi T}>5$ GeV, $p^*_{T}>20$ GeV.
{\it Lower panel:}
solid curve,  $p_{\psi T}>50$ GeV, $p^*_{T}>50$ GeV;
dashed curve, $p_{\psi T}>50$ GeV, $p^*_{T}>20$ GeV;
dotted curve, $p_{\psi T}>20$ GeV, $p^*_{T}>50$ GeV.}
\end{figure}

The  invariant mass distribution in the class of events with minimal $M(\psi c)$ is shown in Fig.~\ref{fig:mass}
with different constraints for the charmonium transverse momentum $p_{\psi T}$ and 
momentum $p_T^*$ of the fragmenting $c$-quark. Calculations were done assuming the collinear factorization scheme. 
Remarkably, the solid and dashed curves corresponding to equal cuts for $p_{\psi T}$,
are not sensitive to the value of $p_T^*$. These curves are practically indistinguishable. 
This observation signals about importance of the fragmentation mechanism. Indeed, in this case
the momentum $p^*_T$ of the fragmenting quark cannot be smaller than $p_{\psi T}$.

Another 
method, similar to the jet clustering algorithm,  
is to select the configuration with the smallest angular separation
between $\psi$ and the accompanying $c$ or $\bar{c}$:
$R=\sqrt{(\Delta\phi)^2+(\Delta\eta)^2}$, where $\Delta\phi$ and $\Delta\eta$ are the difference of azimuthal angles and between of pseudo-rapidities for the produced $c$-quark and $\psi$ respectively. 
This corresponds to the intuitive
picture of correlating comoving products of fragmentation, which are usually considered as a signature of a jet.

\begin{figure}
\epsfig{figure=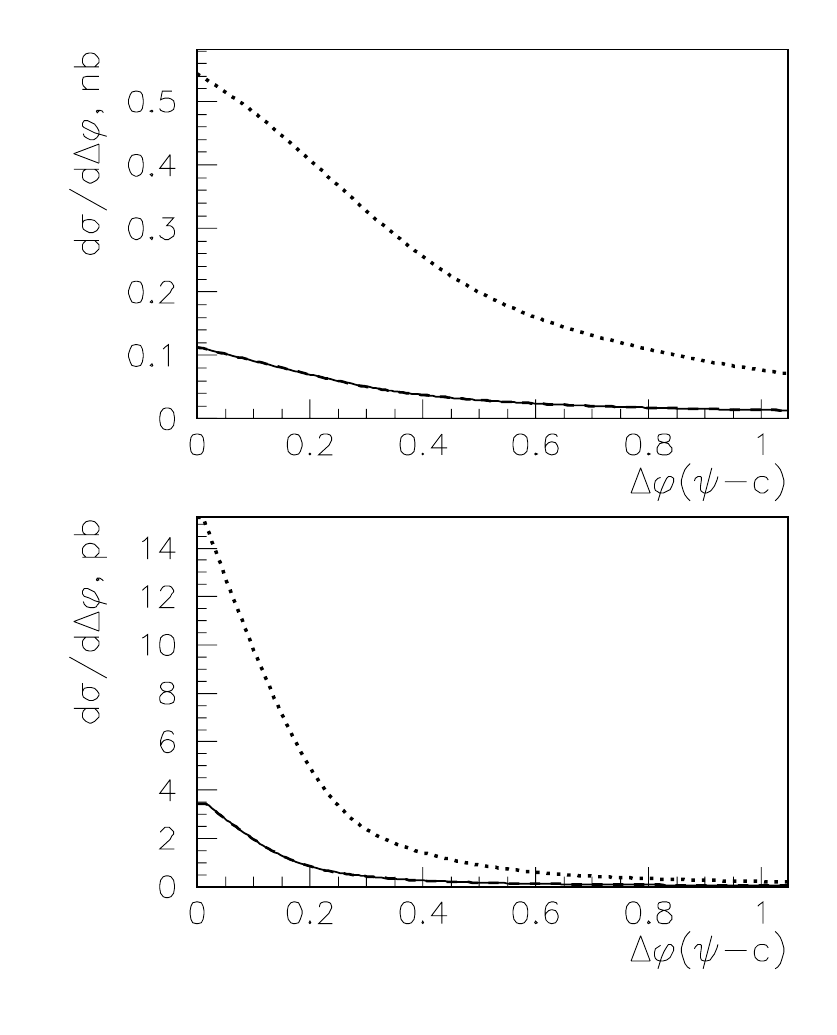,width=8.5cm}\\[-0.8cm]
\caption{\label{fig:dphi} Azimuthal angle difference between the partners in the 
$J/\psi+c$ system under the different kinematic selection rules.
{\it Upper panel:}
solid curve, $p_{\psi T}>20$ GeV, $p^*_{T}>20$ GeV;
dashed curve, $p_{\psi T}>20$ GeV, $p^*_{T}>5$ GeV;
dotted curve, $p_{\psi T}>5$ GeV, $p^*_{T}>20$ GeV.
{\it Lower panel:}
solid curve,  $p_{\psi T}>50$ GeV, $p^*_{T}>50$ GeV;
dashed curve, $p_{\psi T}>50$ GeV, $p^*_{T}>20$ GeV;
dotted curve, $p_{\psi T}>20$ GeV, $p^*_{T}>50$ GeV.}
\end{figure}

\begin{figure}
\epsfig{figure=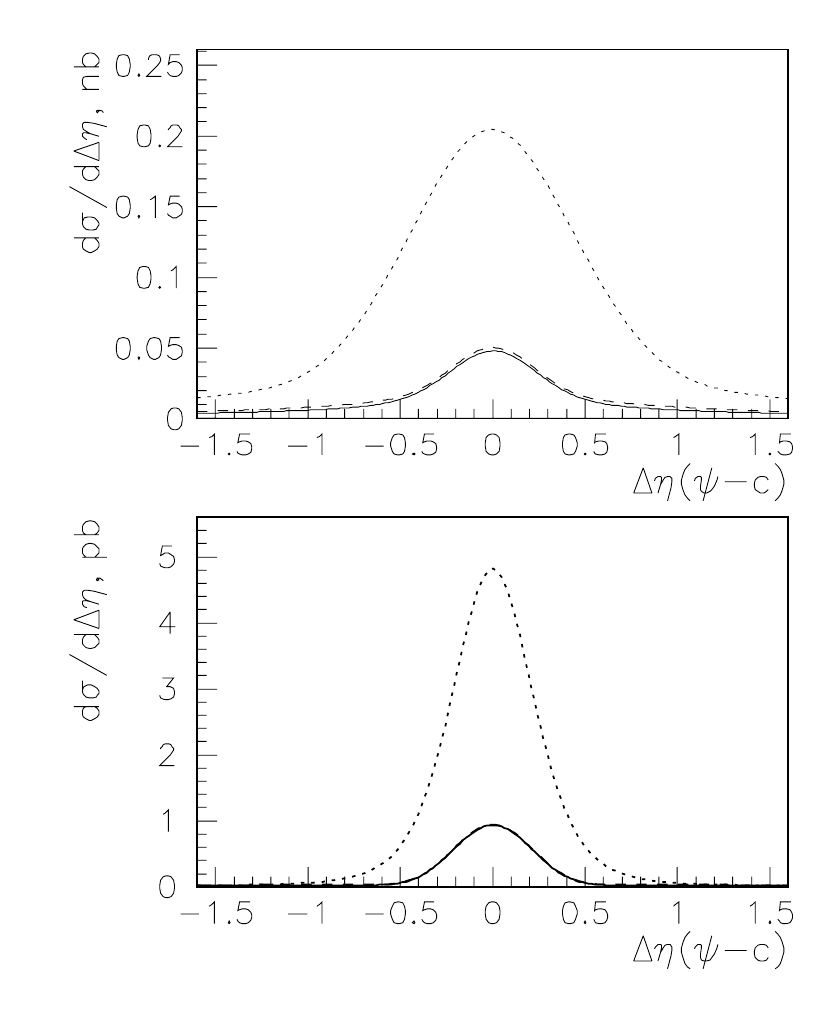,width=8.5cm}\\[-0.8cm]
\caption{\label{fig:deta} Pseudorapidity difference between the partners in the 
$J/\psi+c$ system under the different kinematic selection rules. 
{\it Upper panel:}
solid curve, $p_{\psi T}>20$ GeV, $p^*_{T}>20$ GeV;
dashed curve, $p_{\psi T}>20$ GeV, $p^*_{T}>5$ GeV;
dotted curve, $p_{\psi T}>5$ GeV, $p^*_{T}>20$ GeV.
{\it Lower panel:}
solid curve,  $p_{\psi T}>50$ GeV, $p^*_{T}>50$ GeV;
dashed curve, $p_{\psi T}>50$ GeV, $p^*_{T}>20$ GeV;
dotted curve, $p_{\psi T}>20$ GeV, $p^*_{T}>50$ GeV.}
\end{figure}

We separated all theoretically generated events into two classes $R(c\psi)<R(\bar c\psi)$ and vice versa. 
Then we checked the $\phi$ and $\eta$ correlations in each of these classes of events. Some examples for such distributions, calculated withing the collinear factorization approach, 
are presented in Figs.~\ref{fig:dphi} and \ref{fig:deta} for $\Delta\phi$ and $\Delta\eta$ respectively.

Again, the solid and dashed curves are 
almost indistinguishable: the requirement that $p_{\psi T}$ is large
means automatically that the sum $p_{\psi T}+p_{c T}$ is also large.

We see that with harder cuts on $p_T$'s the system becomes better collimated
(narrower $\Delta\phi$ and $\Delta\eta$ distributions) and so, better suits 
the fragmentation topology. However, the fact that the shape of these 
distributions depends on the selected $p^*_T$ indicates that we are not yet in 
the fragmentation regime.

Remarkably, both methods lead to rather similar consequences.
The quality of our selection rules and the effect of kinematic constraints
on $p_\psi$ and $p^*$ are illustrated in Figs.~\ref{fig:mass} - \ref{fig:deta}.

As far the products of fragmentation can be identified, we can sum up the momenta of the meson and its closest charmed partner and call 
it the parent quark momentum, $p^*=p_\psi+p_c$. The fragmentation variable 
$z$ is then defined in the usual manner: $z=p^+_\psi/p^{*+}$.
Then we can extract the effective fragmentation function $D_{c/\psi}(z)$
and compare it with the fragmentation function from $e^+e^-$.
The fragmentation mechanism is expected to be important at
high transverse momenta, so some kinematic constraints should be imposed.
In Fig.~\ref{fig:Cfragm} we compare the effective fragmentation functions obtained with  different kinematic cuts.
 They disagree with each other and 
both of them disagree with the fragmentation function derived from  $e^+e^-$ 
annihilation. Nevertheless, with a higher cut $p_{\psi T}>50$ GeV, $p^*_{T}>50$ GeV,
the effective fragmentation function at large $z>0.8$ is close to the result from $e^+e^-$.
At the same time, we do not expect any agreement at small $z$, because selecting large values of 
$p_{\psi T}$, we by default suppress production of $\psi$ at small $z$.
 
Notice that while theoretically the parent quark is known, in experiment
its momentum is difficult to reconstruct, which would require reconstruction of the whole jet. 
In inclusive measurements only the momentum of $\psi$ is known,
which introduces ambiguity in identifying the "true fragmentation" region.

In fact, the magnitude of $p_T^*$ cannot be regarded as a decisive signature of the fragmentation mechanism.
Indeed, consider an infinitely small cell $\Delta$ in the quark momentum space 
$d^3p^*$. For every given cell, one can plot a distribution in $z$ normalizing by 
the appropriate cross section of quark pair production,
\begin{eqnarray}
&& \!\!\!\!\!\!\! D(z) = \\ &&\frac{d}{dz}\Bigl[\;
 \!\!\! \int_\Delta \!\! d^3p^*\,\sigma(g\,g\to\psi\,c\,\bar{c})\Bigr] \Bigl/
  \Bigl[\int_\Delta \!\! d^3p^*\,\sigma(g\,g\to c\,\bar{c})\Bigr],\n
\end{eqnarray}
what would give the true quark fragmentation function for a chosen $\Delta(p^*)$.
By definition, the fragmentation function should be independent of the chosen
$p^*$ (provided that $p^*_{T}$ is large enough to fulfil the conditions of the factorization 
theorem). Then, one can average the statistics over many cells,  extending 
the integration in the above formula up to an arbitrarily large part of the phase 
space, $p^*_T>p^*_{T,min}$.
If the distributions calculated for different choices of $p^*_{T,min}$ do not 
coincide, one concludes that the assumption on universality of fragmentation 
function is violated.

The situation in the \ktf case is even more complicated. Here we have an extra
contribution to the transverse momentum that comes from the primordial $k_T$ of the initial
gluons. The large-$k_T$ behavior of the gluon densities is nearly power-like,
${\F}_g(x,k_{T}^2,\mu^2)\sim 1/k_T^4$, and is comparable with the $p_T$ dependence
of the hard scattering matrix element. So, there always presents a non-negligible
contribution to the $p_T$ of essentially non-fragmentation nature.

\begin{figure}
\epsfig{figure=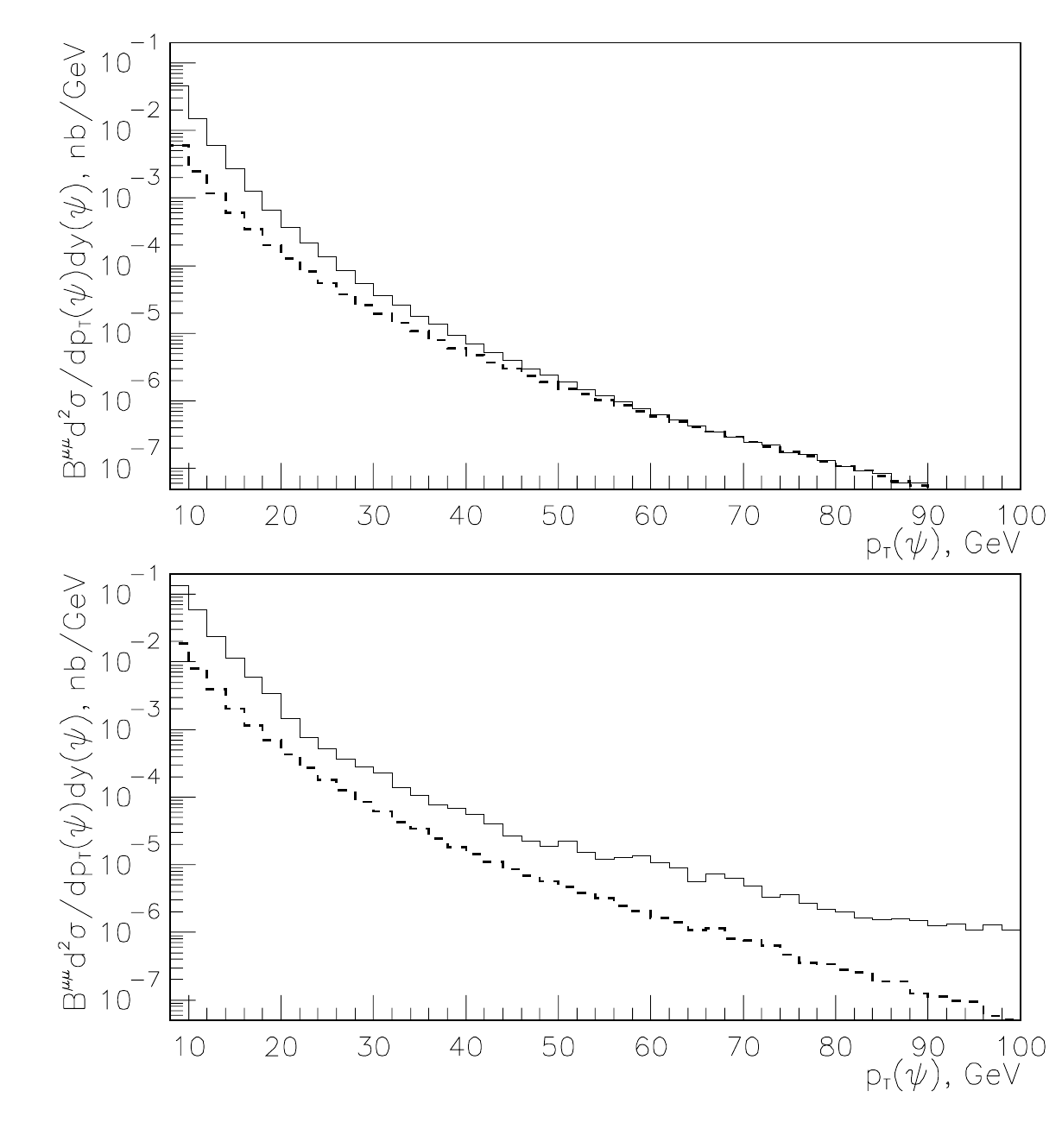,width=8.5cm}\\[-0.5cm]
\caption{\label{fig:pT-dep} Transverse momentum distributions of $J/\psi$ mesons
produced at the mid rapidity in $pp$ collisions at $\sqrt{s}=7$ TeV.
The full $\O (\alpha_s^4)$ calculation 
$g\,g\to\psi\,c\,\bar{c}$ is shown by solid curve. The fragmentation 
approximation, $g\,g\to c\,\bar{c}$, followed by $c\to\psi\,c$ with $D_{c/\psi}(z)$
from $e^+e^-$ annihilation, is presented by dashed curve.
Upper plot is calculated within the collinear factorization scheme with the MSTW gluon densities \cite{MSTW08}; 
lower plot, within the \ktf model with A0 gluon densities \cite{Jung}.}
\end{figure}

These our observations are confirmed by the results of calculations of the $p_T$ spectra of $J/\psi$ mesons
produced in $pp$ collisions at $\sqrt{s}=7$ TeV, depicted in Fig.~\ref{fig:pT-dep}.
The "full LO" and "fragmentation"
curves seem to converge at around $p_{\psi T}\simeq 40$ GeV in the collinear case
(upper panel), but seem to never come together in the \ktf case (lower panel). This figure 
indicates that making use of the fragmentation approach below 40 GeV is by no
means justified. But even at $p_{\psi T}> 40$ GeV the apparent agreement 
between the curves might  be partially a fortune, rather than a consequence of 
the factorization theorem (recall the disagreement between the $z$ distribution
at $p_{\psi T}>50$ GeV in Fig. 6 and the fragmentation function derived from
$e^+e^-$ annihilation).

Going to higher order calculations for the charm fragmentation function
would not help, since the origin of the problem is not in the fragmentation
function on its own, but rather in the unavoidable presence of large non-fragmentation
contributions. Inclusion of the color octet production scheme cannot help either, 
as it would not solve the problem in the color singlet channel and, most 
probably, will suffer from the same troubles, in view of   
much larger number of non-fragmentation diagrams.

\section{Conclusions}

We compared $J/\psi\, c\,\bar c$ production in $pp$ collisions, calculated within the collinear factorization scheme with the full LO set of diagrams,
and the net fragmentation mechanism with the fragmentation function known from $e^+e^-$ annihilation. 
The non-fragmentation contribution is found to be rather large, extending up to
transverse momenta about as high as  40 GeV. These contributions significantly change the slope 
of the $p_{\psi T}$ spectrum in the intermediate region (between 10 and 40 GeV).
The accuracy of the fragmentation approximation cannot  be improved
with either more precise calculations of the charm fragmentation function, or 
including the color octet poduction channels. The presence of essentially
non-fragmentation contributions makes the fragmentation approximation
below 40 GeV groundless.

We also performed calculations within the approach based on the $k_T$-factorization assumption.
In this case the non-fragmentation partial contribution remains important, and even rises at large $p_T$.
This happens because of too large primordial gluon momenta $k_T$, which break down the principle of factorizing  short and long distances in the process.

{\bf Acknowledgments:} 
This work was supported in part by Fondecyt grants 1170319 (Chile), by Proyecto Basal FB 0821 (Chile), and by CONICYT grant  PIA ACT1406 (Chile).

\end{document}